\documentclass[aps, prb, twocolumn, showpacs, superscriptaddress]{revtex4-1}

\usepackage{amsmath,amssymb}
\usepackage{graphicx}
\usepackage{dcolumn}
\usepackage{bm}
\usepackage{color,xcolor}
\usepackage[colorlinks,
            linkcolor=red,
            anchorcolor=blue,
           citecolor=green
            ]{hyperref}
\begin{document}

\title {Periodic Anderson model meets Sachdev-Ye-Kitaev interaction: A solvable playground for heavy fermion physics}
\author{Yin Zhong}
\email{zhongy@lzu.edu.cn}
\affiliation{Center for Interdisciplinary Studies $\&$ Key Laboratory for
Magnetism and Magnetic Materials of the MoE, Lanzhou University, Lanzhou 730000, China}

\date{\today}

\begin{abstract}
The periodic Anderson model is a classic theoretical model for understanding novel physics in heavy fermion systems. Here, we modify it with the Sachdev-Ye-Kitaev interaction, (random all-to-all interaction) thus the resultant model admits an exact solution at large-$N$ (e.g. spin flavor) limit. By analytical field theory arguments and numerical calculations, we establish that the system supports a low-temperature (heavy) Fermi liquid and more interesting a non-Fermi liquid solution at elevated temperature/energy. For physical observable, the latter one contributes a sharp peak at Fermi energy for spectral function, a non-Fermi liquid-like $T^{-1}$ resistivity and shows a robust Fano lineshape in tunneling spectrum. This system may be simulated by ultracold atom gases and can serve as a good playground for studying many ubiquitous symmetry-breaking instabilities like unconventional superconductivity or topological orders in generic heavy fermion systems.
\end{abstract}

\maketitle

\section{Introduction}\label{intr}
Landau's Fermi liquid theory is the cornerstone for our understanding on interacting many-body fermion system above one-spatial dimension.
However, more and more examples of non-Fermi liquid behaviors have been accumulated over recent thirty years after discovery of cuprate high-temperature superconductors.\cite{Lee2006,Rosch2007} One of important battlefields in this direction is the heavy fermion compound, where conduction electron hybridizes with local f-electron to form heavy fermionic quasi-particle and non-Fermi liquid phenomena has been widely observed around the magnetic quantum critical points, e.g. in prototypical f-electron quantum critical materials CeCu$_{6-x}$Au$_{x}$ and YbRh$_{2}$Si$_{2}$.\cite{Rosch2007,Hewson1993,Coleman2015}

A standard theoretical model for heavy fermion physics is the periodic Anderson model (PAM), in which only local f-electron has on-site Hubbard interaction while conduction electron forms a free Fermi sea.\cite{Hewson1993} Although this model is transparent and rather simple in one's mind, analytical and numerical progresses on it have been hindered due to its non-perturbation feature and fermion's minus-sign problem.\cite{Rosch2007,Coleman2015,Jarrell1993,Sugar1995}

On the other hand, an unexpected breakthrough on non-Fermi liquid study has come from the so-called Sachdev-Ye-Kitaev (SYK) model,\cite{Sachdev1993,Kitaev} where random all-to-all interaction between Majorana or complex fermion leads to solvability of such models at large-$N$ limit.\cite{Parcollet1999,Maldacena} At low-energy or strong coupling limit, the fermion Green's function shows power-law dependence on energy without any Fermi liquid quasi-particle weight. Since spatial correlation is not built in the original models, these examples contribute to the local non-Fermi liquid.\cite{Chen2017,Haldar2017a} Further extension to lattice systems gives rise to a crossover from high-temperature local non-Fermi liquid state to low-temperature Landau Fermi liquid state.\cite{Song2017,Jian2017b,Jian2017,Zhang2017,Haldar2017,Patel2017,Chowdhury2018,Dai2018}

Now, inspired by the solvability of SYK model and its resultant novel non-Fermi liquid physics, we propose to study a modified periodic Anderson model with SYK-like interaction, which we call Sachdev-Ye-Kitaev periodic Anderson model (SYK-PAM). The benefit of this model is that it can be solved at large-$N$ limit and both thermodynamic and transport quantities are easy to obtain. We have derived the exact self-consistent equations and found that at strong coupling, this model supports a heavy Fermi liquid solution at low temperature while a non-Fermi liquid regime emerges at high temperature, where the latter one contributes a non-Fermi liquid-like $T^{-1}$ resistivity, has a power-law behavior for susceptibility and shows a Fano lineshape in tunneling spectrum experiments. Numerical results are found to be consistent with analytical arguments and confirm the crossover from low-$T$ to high-$T$ regimes. Furthermore, we have discussed the relationship of our model to realistic heavy fermion compounds and possible experimental realization in ultracold atom gases.

\section{Model}\label{sec2}
The model we study is the SYK-PAM, which reads as follows
\begin{eqnarray}
&&\hat{H}_{SYK-PAM}=-\sum_{ij}\sum_{\alpha}t_{ij}\hat{c}_{i\alpha}^{\dag}\hat{c}_{j\alpha}+V\sum_{j}\sum_{\alpha}(\hat{c}_{j\alpha}^{\dag}\hat{f}_{j\alpha}+\hat{f}_{j\alpha}^{\dag}\hat{c}_{j\alpha})\nonumber\\
&&+E_{f}\sum_{j}\sum_{\alpha}\hat{f}_{j\alpha}^{\dag}\hat{f}_{j\alpha}+\frac{1}{(2N)^{3/2}}\sum_{j}\sum_{\alpha\beta\gamma\delta}U_{\alpha\beta\gamma\delta}^{j}f_{j\alpha}^{\dag}\hat{f}_{j\beta}^{\dag}\hat{f}_{j\gamma}\hat{f}_{j\delta}\label{eq1}
\end{eqnarray}
Here, $t_{ij}$ denotes the hopping integral of conduction electron, $V$ is the hybridization between conduction and local f-electron and $E_{f}$ is the f-electron energy level. The last term is the famous SYK interaction for complex fermion and $U_{\alpha\beta\gamma\delta}^{j}$ is the random all-to-all interaction strength with its Gaussian random mean-value setting to zero ($\overline{U_{\alpha\beta\gamma\delta}^{j}}=0$) and $\overline{(U_{\alpha\beta\gamma\delta}^{j})^{2}}=U^{2}$.\cite{Parcollet1999,Song2017,Chowdhury2018} Note that the sum over spin or flavor index $\alpha,\beta,\gamma,\delta$ is from one to $N$ and it will be setting to infinity to give a sensible large-$N$ limit below.

Before proceeding, it is noted that our model is different from the two-band model studied in Ref.\onlinecite{Chowdhury2018}. In that work, there is no hybridization between conduction and f-electron, instead a random interaction couples conduction electron to f-electron as $\hat{H}_{cf}=\frac{1}{N^{3/2}}\sum_{j}\sum_{\alpha\beta\gamma\delta}V_{\alpha\beta\gamma\delta}\hat{c}_{j\alpha}^{\dag}\hat{f}_{j\beta}^{\dag}\hat{c}_{j\gamma}\hat{f}_{j\delta}$.
Due to this coupling, the conduction electron shows the famous marginal Fermi liquid behavior above a crossover temperature,\cite{Varma1989} which is rather different to ours as can be seen in discussions below.

Then, we turn to the (imaginary-time) path integral formalism of our model, whose partition function is expressed by integrating over the weighted action
\begin{eqnarray}
&&\mathcal{Z}=\int \mathcal{D}c^{\dag}\mathcal{D}c\mathcal{D}f^{\dag}\mathcal{D}fe^{-S_{0}-S_{U}}\nonumber\\
&&S_{0}=\int d\tau\sum_{ij}\sum_{\alpha}c_{i\alpha}^{\dag}(\partial_{\tau}\delta_{ij}-t_{ij})c_{j\alpha}\nonumber\\
&&+V\sum_{j}\sum_{\alpha}(c_{j\alpha}^{\dag}f_{j\alpha}+f_{j\alpha}^{\dag}c_{j\alpha})
+\sum_{j}\sum_{\alpha}f_{j\alpha}^{\dag}(\partial_{\tau}+E_{f})f_{j\alpha}\nonumber\\
&&S_{U}=\int d\tau \frac{1}{(2N)^{\frac{3}{2}}}\sum_{j}\sum_{\alpha\beta\gamma\delta}U_{\alpha\beta\gamma\delta}^{j}f_{j\alpha}^{\dag}f_{j\beta}^{\dag}f_{j\gamma}f_{j\delta}\nonumber
\end{eqnarray}
where the integral of the imaginary time $\tau$ is from zero to $\beta=1/T$ and all fields $c_{j\alpha},f_{j\alpha}$ are the anticommuting Grassman number.

After performing the standard Gaussian random average over each independent $U_{\alpha\beta\gamma\delta}^{j}$ and focusing on one replica realization,\cite{Song2017,Chowdhury2018} we obtain
\begin{eqnarray}
&&\mathcal{Z}=\int \mathcal{D}c^{\dag}\mathcal{D}c\mathcal{D}f^{\dag}\mathcal{D}fe^{-S_{0}-S_{int}}\nonumber\\
&&S_{int}=-\frac{U^{2}}{4N^{3}}\int d\tau\int d\tau' \sum_{j}\sum_{\alpha\beta\gamma\delta}f_{j\alpha}^{\dag}(\tau)f_{j\alpha}(\tau')  f_{j\beta}^{\dag}(\tau)f_{j\beta}(\tau')\nonumber\\
&&\times f_{j\gamma}^{\dag}(\tau')f_{j\gamma}(\tau)f_{j\delta}^{\dag}(\tau')f_{j\delta}(\tau)\nonumber
\end{eqnarray}

\subsection{Effective action and self-consistent equations at large-$N$ limit}
Now, we introduce
\begin{eqnarray}
G_{ij}^{f}(\tau',\tau)=\frac{1}{N}\sum_{\alpha}f_{j\alpha}^{\dag}(\tau)f_{i\alpha}(\tau')\nonumber
\end{eqnarray}
and insert it into the action $S_{int}$ with adding the following constraint term into the partition function,
\begin{eqnarray}
1&&=\int \mathcal{D}G\delta\left(G_{ij}^{f}(\tau',\tau)-\frac{1}{N}\sum_{\alpha}f_{j\alpha}^{\dag}(\tau)f_{i\alpha}(\tau')\right)\nonumber\\
&&=\int \mathcal{D}\Sigma\int \mathcal{D}G e^{\int d\tau\int d\tau'\Sigma_{ji}(\tau,\tau')(NG_{ij}^{f}(\tau',\tau)-\sum_{\alpha}f_{j\alpha}^{\dag}(\tau)f_{i\alpha}(\tau'))}\nonumber
\end{eqnarray}
Therefore, we can rewrite the action $S_{int}$ as
\begin{eqnarray}
S_{int}&&=-\frac{NU^{2}}{4}\int d\tau \int d\tau'\left(G_{jj}^{f}(\tau',\tau)G_{jj}^{f}(\tau,\tau')\right)^{2}\nonumber\\
&&=-\frac{NU^{2}}{4}\int d\tau \int d\tau' |G_{jj}^{f}(\tau',\tau)|^{4}\nonumber
\end{eqnarray}
Here, we have used the fact $G_{ij}^{f}(\tau',\tau)=(G_{ji}^{f}(\tau,\tau'))^{\ast}$ to simplify the above expression. Then, we integrate out all fermions and it leads to
\begin{eqnarray}
&&\mathcal{Z}=\int\mathcal{D}\Sigma\int \mathcal{D}G e^{-S_{eff}}\nonumber\\
&&S_{eff}=-N\int d\tau\int d\tau'\Sigma_{ji}(\tau,\tau')G_{ij}^{f}(\tau',\tau)\nonumber\\
&&-\frac{NU^{2}}{4}\int d\tau \int d\tau' |G_{jj}^{f}(\tau',\tau)|^{4}\nonumber\\
&&-N\mathrm{Tr}\ln[(\partial_{\tau}\delta_{ij}-t_{ij})(\partial_{\tau}\delta_{ij}+E_{f}\delta_{ij}+\Sigma_{ji}(\tau,\tau'))-V^{2}\delta_{ij}]\nonumber\\
&&\label{eq2}
\end{eqnarray}
where $S_{eff}$ is the effective action for $G$ and $\Sigma$. Now, at large-$N$ limit, the partition function is dominated by the extremal $S_{eff}$, which means we can obtain the leading $G$ and $\Sigma$ as
\begin{eqnarray}
&&\frac{\delta S_{eff}}{\delta |G|}=0\Rightarrow\Sigma_{ji}(\tau,\tau')=-U^{2}(G_{jj}^{f}(\tau,\tau'))^{2}G_{jj}^{f}(\tau',\tau)\delta_{ij}\nonumber\\
&&\frac{\delta S_{eff}}{\delta \Sigma}=0\Rightarrow G=\mathrm{Tr}\frac{-1}{\partial_{\tau}\delta_{ij}+E_{f}\delta_{ij}+\Sigma_{ji}-\frac{V^{2}}{\partial_{\tau}\delta_{ij}-t_{ij}}}\nonumber
\end{eqnarray}
Since the system has the translational invariance in both spatial and imaginary-time domain, we can further simplify above equations as
\begin{eqnarray}
&&\Sigma(\tau)=-U^{2}(G^{f}(\tau))^{2}G^{f}(-\tau)\label{eq3}\\
&&G^{f}(k,\omega_{n})=\frac{1}{i\omega_{n}-E_{f}-\Sigma(\omega_{n})-\frac{V^{2}}{i\omega_{n}-\varepsilon_{k}}}\label{eq4}
\end{eqnarray}
where $G^{f}(\tau)\equiv G_{jj}^{f}(\tau)$, $\Sigma(\tau)\equiv\Sigma_{jj}(\tau)$ and $\omega_{n}=\frac{(2n+1)\pi}{\beta}$ is the fermionic
Matsubara frequency. Alternatively, one can obtain these two self-consistent equations by standard Feynman diagrams summation and it is found that $G^{f},\Sigma$ are just the f-electron Green's function and its self-energy, respectively. Similarly, we can derive the Green's function for conduction electron as
\begin{equation}
G^{c}(k,\omega_{n})=\frac{1}{i\omega_{n}-\varepsilon_{k}-\frac{V^{2}}{i\omega_{n}-E_{f}-\Sigma(\omega_{n})}}\label{eq5}
\end{equation}
Therefore, if Eqs.\ref{eq3} and \ref{eq4} have been solved self-consistently, one can obtain the Green's function for both f-electron and conduction electron.

\subsection{Real-frequency self-energy and self-consistent equations}
Before proceeding, we note that most of physical quantities are related to real-frequency retarded Green's function, so it is useful to formalize the self-consistent equations in the real-frequency domain.

Using the analytical continuity $i\omega_{n}\rightarrow\omega+i0^{+}$ and the spectral representation for Green's function, we obtain the real-frequency self-energy as
\begin{eqnarray}
\Sigma(\omega)&&=U^{2}\int d\omega_{1}\int d\omega_{2}\int d\omega_{3} A_{f}(\omega_{1})A_{f}(\omega_{2})A_{f}(\omega_{3}) \nonumber\\ &&\times\frac{f_{F}(\omega_{1})f_{F}(-\omega_{2})f_{F}(-\omega_{3})+f_{F}(-\omega_{1})f_{F}(\omega_{2})f_{F}(\omega_{3})}{\omega+i0^{+}+\omega_{1}-\omega_{2}-\omega_{3}}\nonumber\\
&&\label{eq6}
\end{eqnarray}
At the same time, the retarded Green's function are found to be
\begin{eqnarray}
&&G^{f}(k,\omega)=\frac{1}{\omega-E_{f}-\Sigma(\omega)-\frac{V^{2}}{\omega-\varepsilon_{k}}}\label{eq7}\\
&&G^{c}(k,\omega)=\frac{1}{\omega-\varepsilon_{k}-\frac{V^{2}}{\omega-E_{f}-\Sigma(\omega)}}\label{eq8}
\end{eqnarray}
Here, $\varepsilon_{k}=-\sum_{j}e^{-ikR_{ij}}t_{ij}$ is the dispersion of free conduction electron, $f_{F}(\omega)$ is the usual Fermi distribution function and $A_{f}(\omega)$ is the local density of state (DOS) for f-electron or spectral function of $G^{f}(\omega)$, ($A_{f}(\omega)=-\frac{1}{\pi}\mathrm{Im}G^{f}(\omega)$) which is defined by
\begin{equation}
G^{f}(\omega)\equiv\frac{1}{N_{s}}\sum_{k}G^{f}(k,\omega)=\int d\varepsilon\frac{N(\varepsilon)}{\omega-E_{f}-\Sigma(\omega)-\frac{V^{2}}{\omega-\varepsilon}}\label{eq9}
\end{equation}
with $N(\varepsilon)$ denoting the DOS for the decoupled conduction electron. Now, combining Eqs.\ref{eq6} and \ref{eq9}, it is straightforward to calculate the real-frequency self-energy and the corresponding spectral and Green's function.

It is interesting to note that Eqs.\ref{eq7}, \ref{eq8} and \ref{eq9} are identical to ones in the dynamic mean-field theory (DMFT) for usual PAM model.\cite{Jarrell1995} The reason is that both DMFT and our theory neglect direct spatial correlation between f-electron, thus only local correlation is included and the self-energy of f-electron has no spatial/momentum dependence but is only the function of time/energy.
Additionally, the self-energy formula Eq.\ref{eq6} is similar to its counterpart in the iterative perturbation theory formalism,\cite{Kajueter1996} which is widely used in DMFT calculation.

\section{Analytical results}\label{sec3}
To extract exact Green's function for physically interesting spectral, thermodynamic and transport properties, generally, one has to numerically solve the self-consistent equations (Eqs.\ref{eq3}, \ref{eq4} or Eqs.\ref{eq6}, \ref{eq9}). But, we can gain much qualitative information on the system by inspecting its several limits.

\subsection{Non-interacting limit}
\begin{figure}
\centering
\includegraphics[width=1.10\linewidth]{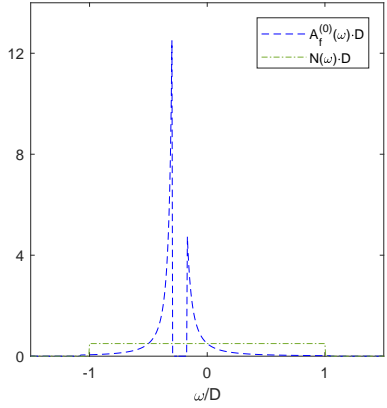}
\caption{\label{fig:dos_free} The DOS of f-electron $A_{f}^{(0)}(\omega)$ at non-interacting limit. $N(\omega)$ is the DOS of the bare conduction electron.}
\end{figure}

When interaction is turned off, we have a non-interacting system with conduction electron hybridizing with local f-electron, whose Hamiltonian reads
\begin{eqnarray}
\hat{H}_{0}=\sum_{k}\sum_{\alpha}[\varepsilon_{k}\hat{c}_{k\alpha}^{\dag}\hat{c}_{k\alpha}+V(\hat{c}_{k\alpha}^{\dag}\hat{f}_{k\alpha}+\hat{f}_{k\alpha}^{\dag}\hat{c}_{k\alpha})+E_{f}\hat{f}_{k\alpha}\hat{f}_{k\alpha}]\nonumber
\end{eqnarray}
Then, the free Green's function are found to be
\begin{eqnarray}
&&G^{c}(k,\omega)=\frac{1}{\omega-\varepsilon_{k}-\frac{V^{2}}{\omega-E_{f}}} \nonumber\\
&&G^{f}(k,\omega)=\frac{1}{\omega-E_{f}-\frac{V^{2}}{\omega-\varepsilon_{k}}} \nonumber
\end{eqnarray}
The pole of these Green's function determines the following quasi-particle spectrum
\begin{equation}
E_{k\pm}=\frac{1}{2}\left[\varepsilon_{k}+E_{f}\pm\sqrt{(\varepsilon_{k}-E_{f})^{2}+4V^{2}}\right]\nonumber
\end{equation}
which gives the usual hybridization energy bands for non-interacting PAM model. If the chemical potential falls into the band gap, the system should be an insulator otherwise we obtain a metallic state. Moreover, note that the free f-electron Green's function can be rewritten as
\begin{eqnarray}
&&G^{f}(k,\omega)=\frac{1}{\omega-E_{f}}\left(1+\frac{V^{2}}{\omega-E_{f}}\frac{1}{\omega-\varepsilon_{k}-\frac{V^{2}}{\omega-E_{f}}}\right) \nonumber\\
&& \nonumber
\end{eqnarray}
which means its local DOS has a simple form
\begin{equation}
A_{f}^{(0)}(\omega)=-\frac{1}{\pi N_{s}}\sum_{k}\mathrm{Im}G^{f}(k,\omega)=\frac{V^{2}}{(\omega-E_{f})^{2}}N(\omega-\frac{V^{2}}{\omega-E_{f}})\label{eq10}
\end{equation}
where $N(\omega)$ is the previously used DOS for free conduction electron. In Fig.\ref{fig:dos_free}, we show an example of $A_{f}^{(0)}(\omega)$, which is calculated from a typical free conduction electron's DOS
\begin{equation}
N(\omega)=\frac{1}{2D}\theta(D-|\omega|)\nonumber
\end{equation}
where $D$ is the half-band-width of the free conduction electron and other model parameters are given as $V/D=1/4,E_{f}/D=-1/4$. Here, it is found that although the DOS for conduction electron is constant over $-D$ to $D$, f-electron has split into two bands with a band gap $\sim V$ around the bare f-electron energy level $E_{f}$. Since Fermi energy is fixed to zero in our model, it is seen that the DOS around Fermi energy is finite and has no singularity. Therefore, if we are only interested in low energy physics around Fermi energy, we can approximate the low energy f-electron's DOS as a constant, i.e $A_{f}^{(0)}(\omega\rightarrow0)\simeq N_{F}\equiv\frac{V^{2}}{E_{f}^{2}}N(\frac{V^{2}}{E_{f}})$.

\subsection{Weak coupling limit}
Next, if the interaction is weak compared to non-interacting bands, we can use perturbation theory to extract physical information reliably. For our model here, the first-order self-energy is obtained by inserting the non-interacting local density of state $A_{f}^{(0)}(\omega)$ into Eq.\ref{eq6}£¬ which reads
\begin{eqnarray}
\Sigma^{(1)}(\omega)&&=U^{2}\int d\omega_{1}\int d\omega_{2}\int d\omega_{3} A_{f}^{(0)}(\omega_{1})A_{f}^{(0)}(\omega_{2})A_{f}^{(0)}(\omega_{3}) \nonumber\\ &&\times\frac{f_{F}(\omega_{1})f_{F}(-\omega_{2})f_{F}(-\omega_{3})+f_{F}(-\omega_{1})f_{F}(\omega_{2})f_{F}(\omega_{3})}{\omega+i0^{+}+\omega_{1}-\omega_{2}-\omega_{3}}\nonumber\\
&&\label{eq11}
\end{eqnarray}

For clarity, we focus on the self-energy at zero-temperature and at the same time use the approximation $A_{f}^{(0)}\simeq N_{F}$, thus we find
\begin{eqnarray}
\Sigma^{(1)}(\omega)&&\simeq U^{2}N_{F}^{3}\int d\omega_{1}\int d\omega_{2}\int d\omega_{3} \nonumber\\ &&\times\frac{\theta(-\omega_{1})\theta(\omega_{2})\theta(\omega_{3})+\theta(\omega_{1})\theta(-\omega_{2})\theta(-\omega_{3})}{\omega+i0^{\dag}+\omega_{1}-\omega_{2}-\omega_{3}}\nonumber
\end{eqnarray}
It is easy to obtain the imaginary part of self-energy as
\begin{equation}
\mathrm{Im}\Sigma^{(1)}(\omega)\simeq-\frac{U^{2}\pi N_{F}^{3}}{2}\omega^{2}+\mathcal{O}(\omega^{4})\propto-\omega^{2}\nonumber
\end{equation}
and the corresponding real part of self-energy is found to be $\mathrm{Re}\Sigma^{(1)}(\omega)\simeq-U^{2}N_{F}^{2}\omega$ by the familiar Kramers-Kronig relation.\cite{Coleman2015} Obviously, the first-order self-energy just tells us that at weak coupling regime, the system should be a Landau's Fermi liquid. Therefore, at weak coupling limit, we may write the f-electron Green's function as
\begin{eqnarray}
G^{f}(k,\omega)=\frac{Z}{\omega-Z\tilde{E}_{f}-\frac{Z\tilde{V}^{2}}{\omega-\varepsilon_{k}}+iaZ\omega^{2}}\label{eq12}
\end{eqnarray}
Here, $Z\sim\frac{1}{1+U^{2}N_{F}^{2}}$ is the quasi-particle weight, $\tilde{E}_{f},\tilde{V}$ are renormalized parameters compared to the bare ones and $a\sim U^{2}N_{F}^{3}$.

\begin{figure}
\flushleft
\includegraphics[width=1.30\linewidth]{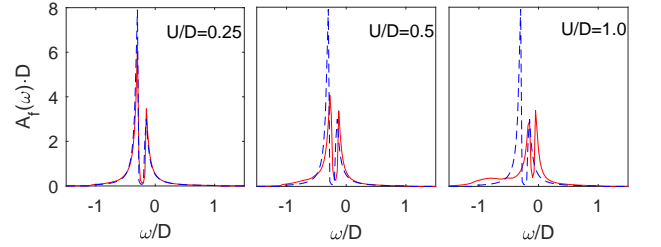}
\caption{\label{fig:dos_weak} The DOS of f-electron from the first-order perturbation theory (Eq.\ref{eq11}) for different $U$, other parameters are the same with Fig.\ref{fig:dos_free}. A small damping factor $\Gamma/D=0.0025$ is used in calculation and $U=0$ case (blue dashed line) is included for comparison.}
\end{figure}

For completeness, we have shown the DOS of f-electron from the first-order perturbation theory (using Eq.\ref{eq11}) in Fig.\ref{fig:dos_weak}. It is found that when the interaction strength $U$ is weaker than the half-band-width of bare conduction electron ($D$), the global behavior of DOS is similar to the non-interacting one and the DOS at Fermi energy ($\omega=0$) is basically unchanged. Another feature is that when increasing interaction strength, the hollow of DOS is shifted toward Fermi energy, which is attributed to the interaction's renormalization effect on the f-electron energy level $E_{f}$.

More generally, one can show the above Fermi liquid-like Green's function (Eq.\ref{eq12}) is indeed the low-energy self-consistent solution of our model by using the exact relation Eq.\ref{eq6}. So, the low-energy physics of our system should be described by conventional Landau's Fermi liquid and this result is not restricted to weak coupling regime.
\subsection{Strong coupling limit}
When interaction $U$ is the largest energy scale ($U\gg D,V,E_{f}$ and here in fact we consider $U\rightarrow\infty$ to simplify the results), we can approximate the f-electron Greens's function as
\begin{eqnarray}
G^{f}(k,\omega)\simeq\frac{1}{-\Sigma(\omega)}\nonumber
\end{eqnarray}
which means the spatial dependence is subleading to the local self-energy and its spectral function is simplified to $A_{f}(\omega)\simeq\frac{1}{-\Sigma(\omega)}$. Then, inserting this spectral function into self-energy equation (Eq.\ref{eq6}), we can find (for $T=0$)
\begin{equation}
G^{f}(\omega)=e^{-i\frac{\pi}{4}}\frac{\pi^{1/4}}{\sqrt{U|\omega|}}\nonumber
\end{equation}
Therefore, the $T=0$ DOS of f-electron at strong coupling limit reads as follows\cite{Parcollet1999}
\begin{equation}
A_{f}(\omega)=\frac{\pi^{-3/4}}{\sqrt{2U}}\frac{1}{\sqrt{|\omega|}}\label{eq13}
\end{equation}
Actually, when the frequency approaches zero, the DOS of f-electron behaves like a sharp $\delta$ function, which emphasize the existence of a zero energy level located at Fermi energy. However, this is just the result of the strong coupling limit and any deviation should lead to a cutoff for this seemingly divergent DOS. In addition, if finite temperature effect is involved, the DOS at finite $T$ is found to be\cite{Parcollet1999}
\begin{eqnarray}
A_{f}(\omega)&&=\frac{1}{\sqrt{UT}}\varphi_{f}\left(\frac{\omega}{T}\right)\nonumber\\
&&=\frac{1}{2\pi^{9/4}\sqrt{UT}}\cosh\left(\frac{\omega}{2T}\right)\left|\Gamma\left(\frac{1}{4}+i\frac{\omega}{2\pi T}\right)\right|^{2}\nonumber
\end{eqnarray}
where we have introduced a scaling function $\varphi_{f}(x)$ and $\Gamma(z)$ is the standard Euler gamma function. Note that $A_{f}(0)=\frac{1}{\sqrt{4UT}}$ is finite.

A careful reader may note that these results are just the standard answers of SYK model. Although our SYK-PAM model has been defined on a lattice and should have spatial correlation, once we move to strong coupling limit, all spatial correlations are quenched by the strong local correlation. Obviously, such SYK solution has no Fermi liquid-like properties, e.g. non-zero quasi-particle weight $Z$ and $\omega^{2}$-dependence in the  imaginary part of self-energy. So, they should be considered as a solution for local non-Fermi liquid with power-law dependence in its spectral function.\cite{Song2017}

Before ending this subsection, we consider the finite but still large $U$ case, where the f-electron Green's function can be approximated as
\begin{eqnarray}
G^{f}(k,\omega)=\frac{1}{\omega-E_{f}-\frac{V^{2}}{\omega-\varepsilon_{k}}+e^{i\frac{\pi}{4}}\sqrt{\frac{2U|\omega|}{\pi^{1/2}}}}\label{eq14}
\end{eqnarray}
From this Green's function, it is clear that the SYK interaction induced non-Fermi liquid self-energy dominates if $|\omega|\gg \frac{E_{f}^{2}}{U},\frac{V^{4}}{D^{2}U}$ or equivalently when $\frac{D^{2}}{U}<|\omega|<U$. (Here $D$ should be considered as the effective band-width of hybridization energy bands of non-interacting PAM model) Therefore, we may define a crossover scale $E^{\ast}\sim \frac{D^{2}}{U}$, above which the local non-Fermi liquid behavior should appear while below this scale, the conventional Landau's Fermi liquid emerges. It is clear that the f-electron Green's function in this Fermi liquid state is also described by Eq.\ref{eq12} and at the present strong coupling regime, we find the quasi-particle weight can be approximated as
\begin{equation}
Z\sim\frac{1}{U^{2}N_{F}^{2}}\ll1 \nonumber
\end{equation}
which means the effective mass is enhanced by $1/Z$ times, thus we finally obtain a heavy Fermi liquid state at strong coupling regime when $|\omega|,T\ll E^{\ast}$.

\subsection{From weak to strong coupling and other physical properties}
From discussions on previous subsections, we have learnt that at weak coupling, the f-electron in our the system forms a conventional Landau's Fermi liquid while at strong coupling limit, we obtain a local non-Fermi liquid state with power-law-like spectral function above the energy scale $E^{\ast}$ and the system returns to a heavy fermion state below $E^{\ast}$.
When combining these two limits, since the ground state is fixed to be (heavy) Fermi liquid, we expect a crossover from Fermi liquid to non-Fermi liquid at elevated temperature/energy regime with increasing of interaction strength.

Previous arguments are focused on the f-electron, here for the conduction electron, we know that its Green's function is
\begin{eqnarray}
G^{c}(k,\omega)&&=\frac{1}{\omega-\varepsilon_{k}-\frac{V^{2}}{\omega-E_{f}-\Sigma(\omega)}}\nonumber\\
&&\simeq\left\{
               \begin{array}{ll}
                 \frac{1}{\omega-\varepsilon_{k}-\frac{ZV^{2}}{\omega-Z\tilde{E}_{f}+iaZ\omega^{2}}}, & \hbox{$U\ll D,T=0$;} \\
                 \frac{1}{\omega-\varepsilon_{k}-\frac{V^{2}}{\omega-E_{f}+e^{i\frac{\pi}{4}}\sqrt{\frac{2U|\omega|}{\pi^{1/2}}}}}, & \hbox{$U\gg D,T=0$.}
               \end{array}
             \right.\nonumber
\end{eqnarray}
Obviously, in the weak coupling regime ($U\ll D$), the conduction electron also forms Fermi liquid. For the strong coupling regime, if the non-Fermi liquid self-energy is dominated, the self-energy of conduction electron is $\sim-iV^{2}(U|\omega|)^{-1/2}$, which is a divergent one and this divergency should be cutoff by the crossover scale $E^{\ast}$. So, we conclude that the conduction electron in the strong coupling regime has non-Fermi liquid behavior above $E^{\ast}$ but remains a Fermi liquid below such scale. We note that the non-Fermi liquid state here is different from the marginal Fermi liquid found in Ref.\onlinecite{Chowdhury2018}, where the self-energy is $\sim-i\omega\ln\frac{U_{f}}{|\omega|}$.

Now, we turn to discuss other physical properties. Firstly, the thermodynamics of the system is determined by the free energy, which at large-N limit is $\mathcal{F}=TS_{eff}$. Here, $S_{eff}$ (Eq.\ref{eq2}) is the effective action calculated by inserting the solution of self-consistent equations. Thus, we find the following expression for free energy $\mathcal{F}$
\begin{eqnarray}
\frac{\mathcal{F}}{N}&&=T\sum_{n,k}\ln[-G_{0}^{c}(k,\omega_{n})]+T\sum_{n,k}\ln[-G^{f}(k,\omega_{n})]\nonumber\\
&&-\frac{3}{4}T\sum_{n}\Sigma(\omega_{n})G^{f}(\omega_{n})\nonumber\\
&&=T\sum_{n}\int d\varepsilon N(\varepsilon)(\ln[i\omega_{n}G_{0}^{c}(\varepsilon,\omega_{n})]+\ln[i\omega_{n}G^{f}(\varepsilon,\omega_{n})])\nonumber\\
&&-\frac{3}{4}T\sum_{n}\Sigma(\omega_{n})G^{f}(\omega_{n})-2T\ln2
\end{eqnarray}
where $G_{0}^{c}(k,\omega_{n})=\frac{1}{i\omega_{n}-\varepsilon_{k}}$ is bare conduction electron Green's function. The summation is over all fermionic Matsubara frequency and $\Sigma(\omega_{n}),G^{f}(\omega_{n})$ can be easily obtained by replacing $\omega$ with $\omega_{n}$ in Eq.\ref{eq6} and \ref{eq9}. At weak coupling, we know the system is a Fermi liquid, thus its free energy should have $\frac{\mathcal{F}}{N}\sim-T^{2}$. For strong coupling, the low-$T$ regime corresponds to a heavy Fermi liquid while high-$T$ regime is a SYK-like non-Fermi liquid, so we expect $\frac{\mathcal{F}}{N}\sim-T^{2}$ for $T\ll E^{\ast}$ and $\frac{\mathcal{F}}{N}\sim-T^{2}-\mathcal{S}_{0}T$ for $T\gg E^{\ast}$ where $\mathcal{S}_{0}$ is the zero-temperature entropy density of SYK-like models.\cite{Maldacena,Song2017}

Next, let us focus on the transport behavior, e.g. the temperature-dependent resistivity $\rho(T)$. At large-$N$ limit, the current-current correlation responsible for electronic transport is dominated by the standard one-loop Feynman diagram of conduction electron, thus the corresponding real part of zero-frequency optical conductance reads\cite{Coleman2015}
\begin{eqnarray}
\mathrm{Re}\sigma^{ij}(0)&&=Ne^{2}\pi\frac{v_{F}^{2}}{3}\delta_{ij}\int_{-\infty}^{\infty}d\varepsilon\int_{-\infty}^{\infty}d\omega N(\varepsilon)A_{c}^{2}(\varepsilon,\omega) \nonumber\\
 &&\times\left[-\frac{\partial f_{F}(\omega)}{\partial\omega}\right]\label{eq16}
\end{eqnarray}
Here, the spectral function of conduction electron $A_{c}(\varepsilon,\omega)$ is
\begin{eqnarray}
A_{c}(\varepsilon,\omega)&&=-\frac{1}{\pi}\mathrm{Im}\frac{1}{\omega-\varepsilon-\frac{V^{2}}{\omega-E_{f}-\Sigma(\omega)}}\nonumber\\
&&\equiv-\frac{1}{\pi}\mathrm{Im}\frac{1}{\omega-\varepsilon-\Sigma_{c}(\omega)}\nonumber
\end{eqnarray}
and $v_{F}$ is the Fermi velocity of bare conduction electron.

When $T>>E^{\ast}$, the conduction electron self-energy $\Sigma_{c}(\omega)$ is dominated by $\Sigma^{-1}$, whose finite temperature behavior is the well-known SYK form $\Sigma(\omega,T)=\sqrt{UT}\Upsilon(\omega/T)$ with $\Upsilon(\omega/T)$ being a non-singular scaling function. Therefore, we obtain $A_{c}(\varepsilon,\omega)\simeq\frac{\sqrt{UT}}{V^{2}}\Upsilon(\omega/T)$
and
\begin{eqnarray}
\mathrm{Re}\sigma(0)&&\simeq Ne^{2}\pi\frac{v_{F}^{2}}{3}\frac{UT}{V^{4}}\int_{-\infty}^{\infty}d\varepsilon N(\varepsilon)\int_{-\infty}^{\infty}d\omega \Upsilon^{2}(\omega/T) \nonumber\\
 &&\times\left[-\frac{\partial f_{F}(\omega)}{\partial\omega}\right]          \nonumber\\
 &&\simeq\frac{n e^{2}}{m}\frac{UT}{V^{4}}\nonumber\\
 &&\propto T
\end{eqnarray}
On the other hand, when $T<<E^{\ast}$, the conduction electron self-energy $\Sigma_{c}(\omega)$ is Fermi liquid-like. At low-$T$, we have the following Drude-like expression
\begin{eqnarray}
\mathrm{Re}\sigma(0)&&\simeq Ne^{2}\pi\frac{v_{F}^{2}}{3}N(0)\int_{-\infty}^{\infty}d\varepsilon A_{c}^{2}(\varepsilon,0)\nonumber\\
&&=Ne^{2}\pi\frac{v_{F}^{2}}{3}N(0)\left(\frac{\mathrm{Im}\Sigma_{c}(0)}{\pi}\right)^{2}\frac{\pi}{-2(\mathrm{Im}\Sigma_{c}(0))^{3}}\nonumber\\
&&=\frac{Ne^{2}v_{F}^{2}N(0)}{3}\frac{1}{-2\mathrm{Im}\Sigma_{c}(0)}\nonumber\\
&&=\frac{ne^{2}\tau_{c}}{m}\nonumber
\end{eqnarray}
where we have defined the lifetime of conduction electron $\tau_{c}$ due to scattering of f-electron degree of freedom
\begin{equation}
\frac{1}{\tau_{c}}\equiv\mathrm{Im}\left[\frac{V^{2}}{E_{f}+\Sigma(0)}\right]=\frac{-V^{2}\mathrm{Im}\Sigma(0)}{(E_{f}+\mathrm{Re}\Sigma(0))^{2}+(\mathrm{Im}\Sigma(0))^{2}}\nonumber
\end{equation}

Thus, we can estimate the $T$-dependent resistivity as
\begin{eqnarray}
\rho(T)=\frac{1}{\mathrm{Re}\sigma(0)}\propto\left\{
               \begin{array}{ll}
                 \frac{V^{2}U^{2}N_{F}^{3}}{\tilde{E}_{f}^{2}}T^{2}, & \hbox{$T<<E^{\ast}$;} \\
                 \frac{V^{4}}{U}T^{-1}, & \hbox{$T>>E^{\ast}$.}
               \end{array}
             \right.\nonumber
\end{eqnarray}
and we observe that the low-$T$ regime shows the typical Fermi liquid $T^{2}$-law while the high-$T$ regime has a non-Fermi liquid-like $T^{-1}$-law.

Finally, the static charge or spin susceptibility $\chi$ can be estimated as
\begin{equation}
\frac{\chi}{N}=\frac{\chi_{c}}{N}+\frac{\chi_{f}}{N}+2\frac{\chi_{cf}}{N}\nonumber
\end{equation}
where $\chi_{c},\chi_{f},\chi_{cf}$ are contributed from conduction electron, f-electron and their mixing. For each susceptibility, one finds
\begin{eqnarray}
\frac{\chi_{\theta}}{N}&&\propto\int_{-\infty}^{\infty}d\varepsilon N(\varepsilon)\int_{-\infty}^{\infty}d\omega_{1} A_{\theta}(\varepsilon,\omega_{1})\int_{-\infty}^{\infty}d\omega_{2} A_{\theta}(\varepsilon,\omega_{2})\nonumber\\
&&\times \left[-\frac{\partial f_{F}(\varepsilon)}{\partial \varepsilon}\right]\nonumber
\end{eqnarray}
At low temperature, we have
\begin{eqnarray}
\frac{\chi_{\theta}}{N}&&\propto N(0)\left[\int_{-\infty}^{\infty}d\omega_{1} A_{\theta}(0,\omega_{1})\right]^{2}\nonumber
\end{eqnarray}
So, if the system is a Fermi liquid, its susceptibility is basically a constant while $\chi_{f}\sim\chi_{cf}\sim T, \chi_{c}\sim T^{3}$ for the non-Fermi liquid regime.

\section{Numerical results}\label{sec4}
In the previous section, we have given several key analytical results at both weak and strong coupling limit. It is argued that the low-energy/temperature regime is a Fermi liquid while we find a local SYK-like non-Fermi liquid at high energy/temperature. Here, we provide some numerical results by solving self-consistent equations (Eqs.\ref{eq6} and \ref{eq9}), which can supplement the analytical arguments.

\subsection{Local DOS of f-electron}
\begin{figure}
\includegraphics[width=1.40\linewidth]{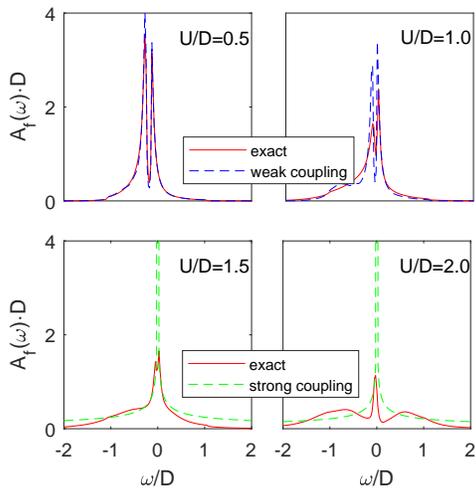}
\caption{\label{fig:dos_tot} The DOS of f-electron from the exact large-$N$ solution for different $U$, other parameters are the same with Fig.\ref{fig:dos_free}. Weak coupling (blue dashed line) and strong coupling (green dashed line) solutions are also included for comparison.}
\end{figure}

In Fig.\ref{fig:dos_tot}, we have shown the exact large-$N$ solution (by solving Eqs.\ref{eq6} and \ref{eq9}) for local DOS of f-electron (red solid line) with different SYK interaction strength $U/D=0.5,1,1.5,2.0$. For comparison, weak (blue dashed line) and strong (green dashed line) coupling solutions are plotted as well.
It can be seen that when interaction is weak ($U<D$), the weak coupling solution is basically indistinguishable from the exact numerical solution.
In contrast, the strong coupling solution successfully captures the main peak structure around Fermi energy, which is confirmed by the exact solution for $U>D$.

In addition, one finds that when increasing interaction from weak to strong coupling regime, the two-peak structure is gradually shifted toward Fermi energy and finally merges into a single peak located at Fermi energy. At the same time, two-hump structure appears when zero-energy peak exists, and their positions are $\pm \frac{U}{2}$. We note that these high-frequency structures are not captured by the strong coupling SYK-like solution, and they should result from the lattice nature of our model, which is beyond the purely local SYK model. In some sense, this two-hump structure is similar to the uppper and lower Hubbard bands, which are familiar ones in the study of Hubbard and PAM models.\cite{Jarrell1995}

\subsection{Temperature-dependent resistivity}
\begin{figure}
\includegraphics[width=1.20\linewidth]{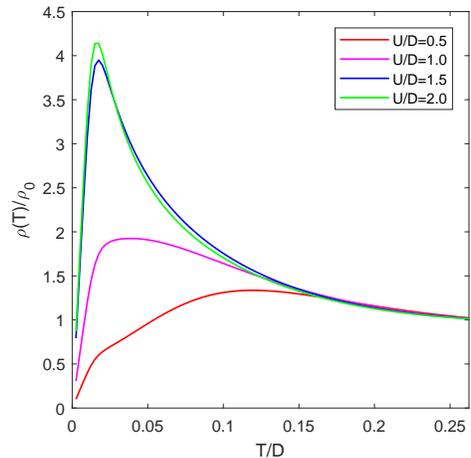}
\caption{\label{fig:rho_T} The temperature-dependent electronic resistivity $\rho(T)$ for different $U$ and $\rho_{0}$ is the high-$T$ resistivity.}
\end{figure}

Finally, we show the temperature-dependent electronic resistivity $\rho(T)$ for different $U$ in Fig.\ref{fig:rho_T}.
For the weak coupling regime, $\rho(T)$ shows the expected Fermi liquid (normal metal) behavior. However, when interaction is further enhanced toward strong coupling,
$\rho(T)$ appears to rise at intermediate temperature regime with non-Fermi liquid-like $T^{-1}$ behavior before returning into low-$T$ Fermi liquid-like regime. This crossover temperature is just
$E^{\ast}\sim \frac{D^{2}}{U}$, where the resistivity reaches its maximum. Therefore, these results agree with our analytical arguments presented in the previous section.

\begin{figure}
\includegraphics[width=1.20\linewidth]{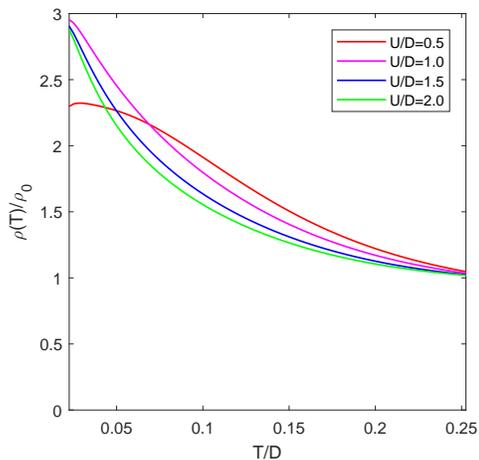}
\caption{\label{fig:rho_s} The temperature-dependent electronic resistivity $\rho(T)$ for SYK-Anderson impurity model. Parameters are the same with Fig.\ref{fig:rho_T}.}
\end{figure}

Actually, the rising of resistivity at high temperature can be ascribed to the single SYK quantum impurity, which is encoded by the following SYK-Anderson impurity model:
\begin{eqnarray}
&&\hat{H}_{SYK-P}=\sum_{k}\sum_{\alpha}\varepsilon_{k}\hat{c}_{k\alpha}^{\dag}\hat{c}_{k\alpha}+V\sum_{k}\sum_{\alpha}(\hat{c}_{k\alpha}^{\dag}\hat{f}_{\alpha}+\hat{f}_{\alpha}^{\dag}\hat{c}_{k\alpha})\nonumber\\
&&+E_{f}\sum_{\alpha}\hat{f}_{\alpha}^{\dag}\hat{f}_{\alpha}+\frac{1}{(2N)^{3/2}}\sum_{\alpha\beta\gamma\delta}U_{\alpha\beta\gamma\delta}^{j}f_{\alpha}^{\dag}\hat{f}_{\beta}^{\dag}\hat{f}_{\gamma}\hat{f}_{\delta}\nonumber
\end{eqnarray}
where conduction electron only couples to single-site f-electron. The corresponding Green's function at large-$N$ limit are found to be
\begin{eqnarray}
&&G^{f}(\omega)=\frac{1}{\omega-E_{f}-\Sigma(\omega)+i\Delta},~~\Delta=\pi N(0)V^{2} \nonumber\\
&&G^{c}(k,\omega)=\frac{1}{\omega-\varepsilon_{k}-V^{2}G^{f}(\omega)}\nonumber
\end{eqnarray}
and their self-consistent equations have similar structure with the SYK-PAM. In Fig.\ref{fig:rho_s}, we have seen that for SYK-Anderson impurity model, its resistivity rises when temperature is lowered and then saturates to a constant due to the constant self-energy $i\Delta$. ($i\Delta$ results from the hybridization of conduction and f-electron $-\sum_{k}\frac{V^{2}}{\omega-\varepsilon_{k}}\simeq i\Delta$.)

\section{Discussion}\label{sec5}

\subsection{Implication for heavy fermion}
Because the random all-to-all SYK interaction in our SKY-PAM model is hardly realized in existing heavy fermion systems, it is not useful to compare with specific heavy fermion compounds. However, we note that the $T$-dependent resistivity found in this work is similar to many heavy fermion metals ($\rho(T)$ rises at high-$T$ and drops at low-$T$ with leaving a visible maximum) though the power-law may be deviated from $T^{-1}$.\cite{Hewson1993,Coleman2015,Rosch2007}

For the spectral function or local DOS, we have found three peak-like structure at strong coupling and these quantities are generally consistent with state-of-art numerical simulations for PAM models, e.g. DMFT and determinant quantum Monte Carlo.\cite{Jarrell1993,Sugar1995} But, we should emphasize that in contrast to the usual PAM model, the main peak centered at Fermi energy does not result from (lattice) Kondo effect, instead it is due to the massive low-energy states in SYK-like models. In heavy fermion experiments, although the DOS is hard to measure directly, one has used the scanning tunneling microscopy (STM) to obtain the spectrum information and visualize the hybridized heavy electron bands.\cite{Schmidt2010,Ernst2011} Theoretically, the STM spectrum $I(\omega)$ is related to the DOS by the following formula\cite{Yang2009,Maltseva2009,Figgins2010,Balatsky2010}
\begin{equation}
I(\omega)\propto t_{f}^{2}A_{f}(\omega)+t_{c}^{2}A_{c}(\omega)+2t_{c}t_{f}A_{cf}(\omega)\label{eq17}
\end{equation}
where $t_{f},t_{c}$ are the STM-tip-electron-tunneling amplitude into f-electron and conduction electron, respectively. Here, $A_{c}(\omega),A_{cf}(\omega)$ are the DOS of conduction electron and its mixing with f-electron. Specifically, we have
\begin{eqnarray}
&&A_{c}(\omega)=\int d\varepsilon N(\varepsilon)A_{c}(\varepsilon,\omega)\nonumber\\
&&A_{cf}(\omega)=\frac{1}{N_{s}}\sum_{k}\left[-\frac{1}{\pi}\mathrm{Im}\frac{V}{\omega-\varepsilon_{k}}G^{f}(k,\omega)\right]\nonumber
\end{eqnarray}
\begin{figure}
\includegraphics[width=1.40\linewidth]{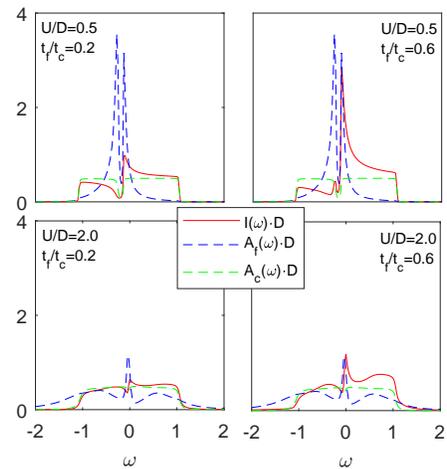}
\caption{\label{fig:stm_figure} The STM spectrum $I(\omega)$ (red line) for weak and strong coupling case. For comparison, the DOS of f-electron $A_{f}(\omega)$ (blue dashed line) and conduction electron $A_{c}(\omega)$ (green dashed line) are also shown.}
\end{figure}
In Fig.\ref{fig:stm_figure}, we have show a typical example for STM spectrum $I(\omega)$ with $t_{c}=1$ and $t_{f}/t_{c}=0.2,0.6$. It is seen that when the tunneling is dominated by conduction electron, (e.g. $t_{f}/t_{c}=0.2$) $I(\omega)$ exhibits the well-known Fano lineshape.\cite{Maltseva2009,Luo2004} The Fano lineshape is widely observed in Anderson impurity systems and also in heavy fermion compound at elevated temperature.\cite{Luo2004,Schmidt2010,Ernst2011} Furthermore, when tunneling into f-electron sites is not small, (e.g. $t_{f}/t_{c}=0.6$)
the lineshape of STM spectrum is still a Fano lineshape for strong coupling case while the weak coupling situation shows two-peak structure like its f-electron DOS. So, the strong coupling regime, which is controlled by the non-Fermi liquid state, seems to show a robust Fano lineshape in the tunneling spectrum experiments.

In addition, we note that the topological Kondo insulator is an interesting and crucial issue in heavy fermion and topological insulator community.\cite{Dzero2016,Dzero2010}
For our model, if we modify the hybridization between conduction and f-electron with the spin-dependent coupling, one can study the correlation effect in such Sachdev-Ye-Kitaev topological Kondo insulator. (Note that a Chern insulator with SYK interaction is discussed in Ref.\onlinecite{Zhang2018}.)
\subsection{Realization of SYK-PAM }
In literature, many authors have proposed that the original SYK model can be realized in ultracold gases, digital quantum simulator, Fu-Kane superconductor, Majorana wires and
graphene flake.\cite{Danshita2017,Solano2017,Chew2017,Pikulin2017,Chen2018} Moreover, some preliminary experimental signals for finite-$N$ SYK model has been found in nuclear magnetic resonance quantum simulator.\cite{Luo2017}

As for the SYK-PAM model, we recall that the usual PAM model is able to be realized in ultracold alkaline-earth metal atom gases.\cite{Nakagawa2015,Zhong2017} So, the main obstacle is to mimic the random all-to-all interaction between fermions with multiple flavors (pseudospin). As argued in Ref.\onlinecite{Danshita2017}, this can be partially solved by coupling two atoms with molecular states via photo-association lasers. Further exploration in this direction is highly desirable and it is promising to see related experiments in near future.

\subsection{Relation to other work}
In Ref.~\cite{Ben-Zion2017}, authors consider a similar model to ours where the hybridization coupling in their model is random and different from ours. They focus on the low energy limit and argue that the resistivity should be $T^{-1/2}$, which qualitatively agrees with our complete analytical and numerical calculation. In our case, we consider all energy regime and give fully consistent numerical solution at large-N limit. Therefore, the high energy details of resistivity and other observable are captured in our work but are lack in their paper. Moreover, the spectrum function and STM spectrum, which are crucial for realistic heavy fermion experiments, are calculated and their basic feature is consistent with underlying SYK physics.

In the other hand, Ref.~\cite{Ben-Zion2017} is motivated by field theory consideration while our work is motivated by searching non-Fermi liquid physics in typical (strongly correlated) condensed matter system (heavy fermion system). So, we have considered relation to heavy fermion experiments and it is found that the global behaviors of resistivity and STM spectrum are similar to experiments.

We also note that a disordered Kondo-Heisenberg model at large-$N$ limit has similar properties to our model.\cite{Burdin} But, the details of physics is different since the Heisenberg interaction is replaced by a Hubbard-like four fermion interaction. So, physical observable like temperature-dependent resistivity has $T^{-1}$ behavior rather than the $T$-linear in that model.

\section{Conclusion and direction for future work}\label{sec6}
In conclusion, we have proposed a modified PAM model with SYK-like interaction and solved it at large-$N$ limit. It is found that this model supports heavy Fermi liquid and non-Fermi liquid states at strong coupling, where the latter one contributes a non-Fermi liquid-like $T^{-1}$ resistivity, has a power-law behavior for susceptibility at high temperature and shows a Fano lineshape in tunneling spectrum experiments. Meanwhile, the spectral function in the non-Fermi liquid state shows three peak-like structure and the main peak locating at Fermi energy reflects the effect of SYK interaction.

In experiments, beside the heavy Fermi liquid and non-Fermi liquid state, superconducting and antiferromagnetic states are ubiquitously encountered in Ce/Yb-based heavy fermion systems when non-Fermi liquid behaviors are suppressed at low-temperature by chemical doping, pressure or external magnetic field. Therefore, it will be interesting to inspect these symmetry-breaking instability in the present model (Note that the instability of the non-Fermi-liquid state in original SYK model has been analyzed in Ref.\onlinecite{Bi2017}, where the time-reversal symmetry is spontaneously broken.) and establishing a global phase diagram will give us further insight into electron correlation effect in both SYK-like models and heavy fermion physics.

\begin{acknowledgments}
This research was supported in part by NSFC under Grant No.~$11704166$ and the Fundamental Research Funds for the Central Universities.
\end{acknowledgments}

\appendix

\end{document}